\documentclass[%
 reprint,
superscriptaddress,
preprintnumbers,
nofootinbib,
 amsmath,amssymb,
 aps,
]{revtex4-1}

\usepackage[colorlinks]{hyperref}
\usepackage{graphicx}
\usepackage{dcolumn}
\usepackage{bm}
\usepackage{color}
\usepackage[dvipsnames]{xcolor}
\usepackage{placeins}
\usepackage{slashed}
\usepackage{multirow}
\usepackage{subfigure}
\usepackage{tabularx}
\usepackage{mathtools}
\usepackage{floatrow}
\usepackage{blindtext}
\usepackage{soul}
\usepackage{slashed}
\usepackage{amsmath}

\hypersetup{
  colorlinks=true,
  linkcolor=red,
  citecolor=blue,
  urlcolor=blue,
  hyperfootnotes=true
}

\definecolor{coral}{RGB}{255,127,80}
\definecolor{indigo}{RGB}{75,0,130}
\definecolor{red}{rgb}{0.9, 0,0}
\definecolor{cerulean}{rgb}{0., 0.62,0.9}
\definecolor{navy}{rgb}{0.05, 0.05,0.8}

\newfloatcommand{capbtabbox}{table}[][\FBwidth]

\newcommand{\GeV}{{\rm GeV}}
\newcommand{\MeV}{{\rm MeV}}
\newcommand{\keV}{{\rm keV}}

\begin{document}

\preprint{TTP22-058, P3H-22-095}
\title{Axion Dark Matter from Lepton flavor-violating Decays}
\author{Paolo Panci}
\affiliation{Dipartimento di Fisica E. Fermi, Universit\`a di Pisa, Largo B. Pontecorvo 3, I-56127 Pisa, Italy }
\affiliation{INFN, Sezione di Pisa, Largo Bruno Pontecorvo 3, I-56127 Pisa, Italy}
\author{Diego Redigolo}
\affiliation{INFN, Sezione di Firenze, Via G. Sansone 1, 50019 Sesto Fiorentino, Italy}
\author{Thomas Schwetz}
\affiliation{Institut f\"ur Astroteilchenphysik, Karlsruhe Institute of Technology,
 Karlsruhe, Germany}
\author{Robert Ziegler}
\affiliation{Institut f\"ur Theoretische Teilchenphysik, Karlsruhe Institute of Technology, Karlsruhe, Germany}

\date{\today}

\begin{abstract}
We propose simple scenarios where lepton flavor-violating couplings generate the observed dark matter abundance through freeze-in of an axion-like particle with mass in the few keV range. Compared to flavor-diagonal freeze-in, this mechanism enhances dark matter stability, softens stellar cooling constraints and improves the experimental sensitivity of accelerator-based searches. These scenarios can be tested by future X-ray telescopes, and in some cases will be almost entirely probed by new searches for lepton flavor violation at high-intensity experiments such as Mu3e and MEG II.
\end{abstract}

\maketitle

\section{Introduction}
Light axions with tree-level flavor-violating couplings to Standard Model (SM) fermions allow to test enormously large scales of Peccei-Quinn breaking at high-intensity laboratory experiments~\cite{MartinCamalich:2020dfe, Calibbi:2020jvd}. Scales up to about $f_a \sim 10^{12}\ \GeV$ are probed by NA62 in  $s-d$ transitions~\cite{NA62:2021zjw, Goudzovski:2022vbt}, while lepton flavor-violating (LFV) muon decays give sensitivity to scales up to $f_a \sim 10^{10} \ \GeV$ \cite{Calibbi:2020jvd, Jho:2022snj}. Such large decay constants imply that the axion can be stable on cosmological scales, motivating scenarios where the axion itself fully accounts for the Dark Matter (DM) relic density today.

 In general, flavor-violating axion couplings depend on the misaligment of PQ charges and SM Yukawas, which can be only fixed if  a theory of flavor is specified~\cite{Ema:2016ops, Calibbi:2016hwq, Linster:2018avp}. Here instead we present a scenario where LFV decays of SM leptons are directly responsible for producing axion DM in the early Universe through thermal freeze-in, so that the DM abundance is set by the product of axion mass and LFV decay rate, $\Omega_{\rm DM} \propto m_a \Gamma (\ell_i \to \ell_j a)$.  The viable range of axion masses is thus bounded from above by the kinematic threshold and from below by direct searches for the flavor-violating decay. This setup gives rise to a class of very simple and predictive models of axion DM, which can be tested with future X-ray~and low energy $\gamma$-ray telescopes \cite{XQC:2015mwy,Neronov:2015kca,Thorpe-Morgan:2020rwc,Ando:2021fhj,Coogan:2021rez,Dekker:2021bos}, and in some cases will be almost entirely probed in future LFV experiments at MEG II and Mu3e~\cite{Calibbi:2020jvd, Jho:2022snj,Perrevoort:2018ttp}, if new search strategies for light particles will be implemented.   
 
 Freeze-in production of axions in the early Universe has been considered before in Ref.~\cite{Baumann:2016wac, DEramo:2018vss, DEramo:2021usm} in order to constrain flavor-violating couplings of the QCD axion from Dark Radiation. Here instead the LFV decays account for  the total observed amount of axion DM. 
\section{The framework}
We consider a ``lepto-philic'' anomaly-free axion, which is a massive pseudo-Goldstone boson $a$
that only couples to SM leptons according to the effective Lagrangian 
\begin{align}
{\cal L}_{\rm eff} = \frac{\partial_\mu a}{2 f_a} \overline{f}_i \gamma^\mu \left( C^V_{f_i f_j} + C^A_{f_i f_j} \gamma_5 \right) f_j - \frac{m_a^2}{2} a^2\, ,
\label{Lag} 
\end{align}
where $C^{A,V}_{f_i f_j}$ are traceless hermitian matrices in lepton flavor space $f = \ell, \nu$. They originate from rotating the charge matrices of the underlying, spontaneously broken $U(1)_X$ symmetry to the mass basis:
\begin{align}
C^{V,A}_{\ell_i \ell_j}  & =  V^\dagger_{R}X_{R} V_{R} \pm  V^\dagger_{L}X_{L} V_{L} \, , & C^{V,A}_{\nu_i \nu_j}  & =  \pm  V^\dagger_{\nu} X_{L} V_{\nu} \, , 
\end{align}
where $X_{R}$ ($X_{L}$) are the traceless $U(1)_X$ charges of $SU(2)_L$ singlet (doublet) fields, and the unitary matrices are defined by $V_{L}^\dagger M_e V_{R}  = M_e^{\rm diag}$ , $V_{\nu}^T M_\nu V_{\nu}  = M_\nu^{\rm diag}$.  Left-handed charged lepton and neutrino rotations are related by the PMNS matrix $V_{\rm PMNS}  = V_{L}^\dagger V_\nu $.

In the following we will discuss two simple scenarios: first we consider the case where only right-handed (RH)  leptons of first and second generation are charged under $U(1)_X$, so that without loss of generality $X_R = {\rm diag} (1,-1,0)$ and $X_L = 0$. We refer to this case as ``two-flavor scenario''. In the second scenario we consider general traceless charges for left-handed (LH) fields  $X_L = {\rm diag} (1,X,-1-X)$ and $X_R = 0$.
In the two-flavor case the rotation matrix is taken to be a general rotation in the 1-2 space parameterized by an angle $ 0 \le \alpha \le \pi/2$, suitably defined such that in the mass basis 
\begin{align}\label{eq:rh-scenario}
C^{V}_{\ell_i \ell_j} & = C^{A}_{\ell_i \ell_j}  = \begin{pmatrix} s_{ \alpha} &c_{ \alpha}  & 0 \\ c_{ \alpha}  &-s_{ \alpha} &0  \\ 0 & 0 & 0 \end{pmatrix}   \, , &
C^{V,A}_{\nu_i \nu_j} & =  0 \, .
\end{align}
We denote this case as the ``$\mu e$-scenario"; analogously we consider also the ``$\tau \mu$-scenario" and the ``$\tau e$-scenario". 
In the second case, dubbed the ``PMNS-scenario", we will take the rotation matrix in the LH sector to be the PMNS matrix, so that
$V_{\rm PMNS} \approx V_L^\dagger$ and
$V_{\nu}$ is close to the identity. This gives in the mass basis
\begin{align}\label{eq:lh-scenario}
C^{V}_{\ell_i \ell_j} & = - C^{A}_{\ell_i \ell_j}  = V_{\rm PMNS} \,{\rm diag} (1,X,-1-X) V_{\rm PMNS}^\dagger\, , \nonumber \\
C^{V}_{\nu_i \nu_j} & = - C^{A}_{\nu_i \nu_j}  =  {\rm diag} (1,X,-1-X)  \, .
\end{align}
Both scenarios  depend on three free parameters: $f_a$, $m_a$ and $\alpha$ ($X$) in the first (second) case, respectively. One combination of these parameters is fixed to reproduce the observed DM relic abundance, leaving a 2-dimensional parameter space in each scenario. 
\subsection{Dark Matter Stability}
In order to be stable on cosmological scales, the axion must be sufficiently light such that the decay channel into electrons is kinematically closed. In the following, we will therefore consider $m_a \ll 1 \text{ MeV}$, so that the axion can only decay into photons and possibly neutrinos with a total decay rate $\Gamma_a =  \Gamma_{a \to \gamma \gamma} + \Gamma_{a \to \nu \nu} \equiv 1/\tau_a^{\gamma\gamma} + 1/\tau_a^{\nu\nu}$, where
\begin{align}
  \displaystyle\Gamma_{a \to \gamma \gamma} &\approx \frac{\alpha_{\rm em}^2}{64 \pi^3} \frac{m_a^3}{f_a^2}  \left| \sum_i C^A_{\ell_i \ell_i} \frac{m_a^2}{12 m_{\ell_i}^2}\right|^2 \, , 
  \end{align}
  up to higher powers of $m_a^2/m_{\ell_i}^2$, and 
  \begin{align}
  \Gamma_{a \to \nu \nu} &\approx \frac{m_a}{8 \pi} \sum_i \left|C^A_{\nu_i \nu_i} \frac{m_{\nu_i}}{f_a}\right|^2\ .  
\end{align}
The decay rate into photons is  suppressed by $m_a^4/m_\ell^4$, since there is no electromagnetic anomaly (cf. Ref.~\cite{Nakayama:2014cza,Takahashi:2020bpq, Han:2020dwo, Han:2022iig, Sakurai:2022roq}), while the decay into neutrinos is suppressed by the smallness of neutrino masses.  The inverse partial widths into neutrinos and photons  read 
\begin{align}
\tau_a^{\nu\nu} & \approx\displaystyle    10^{20} \sec  \left( \frac{60 \, \keV}{m_a} \right) \left( \frac{(0.05\text{ eV})^2}{ \sum_i ( C^A_{\nu_i\nu_i} m_{\nu_i})^2 } \right)     \left( \frac{f_a }{10^{9} \GeV} \right)^{\hspace{-0.05cm} 2} \, , \nonumber  \\
\tau_a^{\gamma\gamma} & \approx\displaystyle    10^{20} \sec  \left( \frac{60 \, \keV}{m_a} \right)^{7}  \left( \frac{m_{\ell_i}}{m_e} \right)^{ 4}  \left( \frac{f_a/C^A_{\ell_i\ell_i} }{10^{9} \GeV} \right)^{2}  \, ,
\label{eq:lifetime}
\end{align}
where $C^A_{\ell_i\ell_i}$ is the axion coupling to the lighter SM lepton. To good approximation $\Gamma_{a \to \gamma \gamma}$ is dominated by the electron loop for generic diagonal couplings to leptons. Since the ratio $\Gamma_{a \to \gamma \gamma}/ \Gamma_{a \to \nu \nu}$ strongly depends on the axion DM mass ($\propto m_a^6$),  for a fairly light ALP, the main decay channel is expected to be into neutrinos.  

For axion decay constants in the right ballpark for producing DM through freeze-in of LFV decays (see next section), the axion lifetime exceeds the age of the Universe by roughly three orders of magnitude, making it possible for the axion to be DM. 

 The axion partial width into photons is further constrained by 
precise measurements of the CMB temperature and polarization anisotropies, which give a bound on the lifetime of roughly $\tau_a^{\gamma\gamma}\gtrsim3\times 10^{24}\text{ sec}$ in the mass range of interest~\cite{Slatyer:2016qyl, Bolliet:2020ofj}. Stronger constraints can be derived from indirect searches for X-rays:
\begin{equation}\label{eq:Xray_bound}
  \tau_a^{\gamma\gamma} \gtrsim (10^{26} \to 10^{28}) \, \text{sec} \,, 
\end{equation}
depending on the precise mass range~\cite{Boyarsky:2005us,Boyarsky:2006fg,Boyarsky:2006zi,Boyarsky:2006kc,Boyarsky:2006hr,Watson:2006qb,Boyarsky:2006ag,Yuksel:2007xh,Boyarsky:2007ay,Loewenstein:2012px,Urban:2014yda,Tamura:2014mta,Ruchayskiy:2015onc,Riemer-Sorensen:2009zil,Riemer-Sorensen:2009zil,Watson:2011dw,Essig:2013goa,XQC:2015mwy,Perez:2016tcq,Roach:2019ctw,Ng:2019gch,Laha:2020ivk,Siegert:2021upf,Roach:2022lgo,Foster:2022ajl}
with a weak dependence on the DM density profile~\cite{Cirelli:2012ut, Laha:2020ivk}.
We collected in Appendix A the relevant constraints for decaying DM in the keV-MeV range in Fig.~\ref{fig:decayDM}, together with the projected sensitivities of future telescopes~\cite{XQC:2015mwy,Neronov:2015kca,Thorpe-Morgan:2020rwc,Ando:2021fhj,Coogan:2021rez,Dekker:2021bos}. 

Axion DM decaying into neutrino-antineutrino pairs will lead to lines with energy below $1\, \MeV$~\cite{Garcia-Cely:2017oco, Lin:2022xbu}, well below the kinematic threshold of the inverse $\beta$ decay reaction $\bar{\nu}_e+ p\to n + e^+$ which requires $E_{\bar{\nu}_e}>1.8\text{ MeV}$. Elastic scattering with electrons is the only active process for these low energy neutrinos which makes it very challenging for present detectors such as Borexino~\cite{Borexino:2010zht} to provide any interesting constraints on decaying DM beyond the cosmological bound. The situation could be possibly improved by future experiments looking for neutrino capture on beta decaying nuclei~\cite{Weinberg:1962zza, Cocco:2007za, McKeen:2018xyz, Chacko:2018uke}.

\subsection{Dark Matter Production}\label{sec:production}

For values of the axion decay constant allowed by the constraints on decaying DM, the axion was never in thermal equilibrium with the SM in the early Universe. Therefore the different contributions to the DM abundance come from freeze-in of LFV decays of SM leptons and leptonic $2\to 2$ scattering processes, and possibly from non-thermal production mechanisms.

Thermal axions will be dominantly produced through their couplings to SM leptons via the freeze-in mechanism~\cite{Hall:2009bx}. Assuming that axion production happens during radiation domination and that the effective number of relativistic degrees of freedom in the SM bath is approximately constant, one can derive a simple analytic expression for the axion relic density, see e.g. Ref.~\cite{DEramo:2020gpr}. For flavor-diagonal couplings, axion DM is produced through the scattering processes $\ell \ell \to \gamma a$ and $\ell \gamma \to   \ell a$ with the abundance roughly given by 
\begin{widetext}
\begin{align}
\Omega_a h^2|_{\text{scattering}} \approx 9\times 10^{-4}\left(\frac{m_a}{50\text{ keV}}\right)\left(\frac{  5\times10^9 \GeV}{f_a/C_{\ell \ell}}\right)^2\left(\frac{m_{\ell
}}{m_\tau}\right)\left(\frac{75}{g_*(m_{\ell})}\right)^{3/2}\ .
\end{align}
\end{widetext}

Flavor-violating axion couplings instead induce freeze-in production through lepton decays, which depend on the FV decay rate  $\Gamma (\ell_i \to \ell_j a) \approx C_{\ell_i \ell_j}^2m_{\ell_i}^3/64\pi f_a^2$. These decays dominate over the flavor-diagonal scattering processes for $C_{\ell_i \ell_j}\sim C_{\ell_i \ell_i}$, and lead to an axion abundance given by
\begin{widetext}
\begin{align}
\Omega_a h^2|_{\ell_i \to \ell_j a} \approx  
 0.12   \left( \frac{m_a}{50 \, \keV} \right)\left(\frac{m_{\ell_i}}{m_\tau}\right) \left( \frac{  5\times10^9 \GeV}{f_a/C_{\ell_i \ell_j}} \right)^2\left(\frac{75}{g_*(m_{\ell_i})}\right)^{3/2}  \, , 
\label{eq:relic}
\end{align}
\end{widetext}
with $C_{\ell_i \ell_j} \equiv \sqrt{|C_{\ell_i \ell_j}^V|^2 + |C_{\ell_i \ell_j}^A|^2 }$.
Remarkably, the required decay constants to account for the total DM relic abundance today are in the range $f_a \sim 10^9-10^{10}\, \GeV$,  which is allowed by present stellar cooling constraints~\cite{MillerBertolami:2014rka, Calibbi:2020jvd} and within the reach of future experiments hunting for LFV in muon decays ~\cite{Calibbi:2020jvd, Jho:2022snj,Perrevoort:2018ttp}. For the same range of decay constants the freeze-in of scattering processes give a negligible contribution to the relic abundance, as a result of the $\alpha_{\text{em}}$ suppression of the scattering rate compared to the decay rate. Since flavor-violating couplings also help to suppress axion decays to photons (which are subject to very stringent constraints as discussed above), they play a crucial role in both production and stability of axion DM (see also Ref.~\cite{Jaeckel:2013uva}).  We finally note that the freeze-in from muon decays is only mildly suppressed compared to the one from tau decays, because the mass hierarchy between the two SM leptons is almost entirely compensated by the large difference in the number of relativistic degrees of freedom in the SM bath.  
 
Freeze-in scenarios are potentially sensitive to DM production processes that are dominated by high temperatures. In the present case, such processes arise from $2\to2$ scattering via the dimension-5 operator of the form 
\begin{align}
{\cal L}_{\rm eff} = - \frac{ia}{f_a} \frac{m_{\ell_i}}{v} C^A_{\ell_i \ell_j} h \overline{\ell}_{i} P_R \ell_{j} + {\rm h.c.}\ , 
\end{align}
which is obtained from Eq.~\eqref{Lag} upon integrating by parts and using the fermion equation of motions (or equivalently performing $a$-dependent fermion field redefinitions), and for simplicity we have set $C^V_{\ell_i \ell_j} = C^A_{\ell_i \ell_j}$. This interaction contributes to freeze-in production of axions via $\ell_i h \to \ell_j a$ and similar processes, which are UV-sensitive, i.e.\ they depend on the reheating temperature $T_R$ (cf. Ref.~\cite{Hall:2009bx}), giving an abundance today 
\begin{align}
\Omega_{ a} h^2\vert_{\text{UV}} & \simeq  \frac{m_{\ell_i} T_R} {3 \pi^3 v^2}\times \Omega_a h^2|_{ \ell_i \to \ell_j  a}  \, .
\end{align}
Requiring the UV contribution to be subdominant with respect to production from LFV decays then gives an upper bound on the reheating temperature
$T_R^{\text{max}}\simeq 3\pi^3 v^2/m_{\ell_i} \approx 2 \times 10^6 \text{ GeV}\times (m_\tau/m_{\ell_i})$, which can be expressed as an upper bound on the Hubble scale at reheating
\begin{equation}\label{eq:IRdominated}
H_R \ll  4 \text{ keV} \left(\frac{m_\tau}{m_{\ell_i}}\right)^2  \,, 
\end{equation}
where we used $g_*(T_R^{\text{max}}) =  106.75$ since $T_R^{\text{max}} \gg v$. For muon decays this gives an upper bound $H_R <1$~MeV.

With this requirement in mind, we can now discuss non-thermal production mechanisms. Potentially most relevant is the misalignment mechanism for ALPs~\cite{Arias:2012az,Blinov:2019rhb}, which generalizes the classic production mechanism for the QCD axion~\cite{AxionDM1,AxionDM2,AxionDM3}. In particular, given the upper bound on the Hubble scale during the radiation domination epoch in Eq.~\eqref{eq:IRdominated}, the onset of axion oscillations (roughly\footnote{For the numerical values below we have used $m_a = 1.6 H(T_{\rm osc})$ as suggested in Ref.~\cite{Blinov:2019rhb}.} defined by the equation $m_a\simeq H$) happens prior to reheating in the mass range of interest (unless for muon decays the reheating temperature saturates the upper limit in Eq.~\eqref{eq:IRdominated}, $H_R \sim 1 \, \MeV$). The present day abundance generated by misalignment is then suppressed due to the dilution that occurred during the initial period of matter domination between the onset of oscillations and $T_R$. The resulting abundance is independent of the ALP mass and reads~\cite{Visinelli:2009kt, Blinov:2019rhb, Arias:2021rer} 
\begin{align}
\Omega_{a} h^2\vert_{\text{mis.}}\! \approx\! 1.2\times 10^{-3}\! \left( \frac{f_a \theta_0}{10^{10}\,  \GeV} \right)^2\!\!\left(\frac{H_R}{1\, \keV}\right)^{1/2}  \,. \label{eq:mis}
\end{align}
Given the upper bound on the reheating temperature necessary to have an IR dominated freeze-in, the misalignment contribution turns out to be always subleading, although in the case of muon LFV decays one can arrange for a sizable misalignment contribution for reheating temperatures close to the upper limit of Eq.~\eqref{eq:IRdominated}. The misalignment contribution could also be enhanced if instead of a period of matter domination inflation ends in a period of kination~\cite{Visinelli:2009kt, Blinov:2019rhb, Arias:2021rer}.

Another available non-thermal production mechanism goes through inflationary perturbations~\cite{Ford:1986sy,Chung:1998zb,Graham:2018jyp,Redi:2022zkt}. This depends on the Hubble scale during inflation $H_I$, which in general can be much higher than $H_R$. In the case of instantaneous reheating, $H_I\sim H_R$ is smaller than the ALP mass in the range of interest and the inflationary production is exponentially suppressed. For $f_a>H_I>m_a$ the global symmetry associated with the axion is always broken during inflation, and quantum fluctuations of the axion field of order $\theta(x)\sim H_I/2\pi f_a$ are generated together with the axion zero-mode $\theta_0$. These fluctuations will generate isocurvature perturbations in the matter power spectrum, which are constrained by CMB observation to be less than a percent of the adiabatic ones. If $H_I\gg H_R$ and we keep requiring $H_R<m_a$, Eq.~\eqref{eq:mis} ensures that the energy density into isocurvature perturbation is small enough as long as $H_I< f_a$. For $H_I>f_a$, the global symmetry is restored during inflation and the axion inflationary production becomes dependent on the radial mode dynamics.

In principle one could relax the constraint in Eq.~\eqref{eq:IRdominated} and consider axion DM scenarios where the abundance receives contributions also from UV dominated freeze-in, misalignment or inflationary production. In what follows we instead assume that $f_a > H_I > m_a > H_R$, so that the dominant production of axion dark matter always comes from   freeze-in of LFV decays and is insensitive to UV dynamics. We will show how this  simple scenario leads to concrete targets for future LFV experiments~\cite{Calibbi:2020jvd, Jho:2022snj,Perrevoort:2018ttp}.

\subsection{Warm Dark Matter}

Even if it is never in equilibrium with the SM bath, axion dark matter produced through freeze-in of LFV decays is created with a large initial velocity and is initially free-streaming. This feature leaves its footprint on the DM power spectrum suppressing the growth of primordial fluctuations with wavelength larger than the DM free-streaming length roughly estimated as  $\lambda_{\text{f.s.}}\simeq 0.1\text{ Mpc}\times (m_a/1\text{ keV})$. The suppression of the power spectrum at small scales can be tested through the absorption features of the spectra of distant quasars generated by neutral hydrogen filaments which are assumed to trace the matter power spectrum~\cite{Viel:2004bf} (denoted by ``Ly$\alpha$" below). Other possible probes involve counting dwarf satellite galaxies of the Milky Way or other nearby galaxies, but are typically more easily affected by observational and astrophysical uncertainties. For this reason we focus on Ly$\alpha$ constraints in this discussion, whose robustness against astrophysical modelling and observational uncertainties has been thoroughly investigated in Ref.~\cite{Boyarsky:2008xj}. A stringent lower bound on the warm DM mass $m_{\rm WDM}^{\rm min} \approx 5.3 \, \keV$ has been derived in Refs.~\cite{Viel:2013fqw,Baur:2015jsy,Irsic:2017ixq}, which can be relaxed to  $m_{\rm WDM}^{\rm min} \approx 3.5 \, \keV$ under more conservative assumptions. Recent works recasted these limits by computing the exact DM distribution function for different freeze-in processes~\cite{Ballesteros:2020adh,DEramo:2020gpr,Decant:2021mhj} . For LFV axion freeze-in this constraint can be written as  
\begin{equation}
m_a\gtrsim 10\text{ keV}\left(\frac{m^{\rm min}_{\text{WDM}}}{3.5\text{ keV}}\right)^{4/3}\left(\frac{75}{g_*(m_\ell)}\right)^{1/3} \, .\label{eq:WDM}
\end{equation}

\begin{figure*}[t]
\begin{center}
	\includegraphics[width=0.482\textwidth]{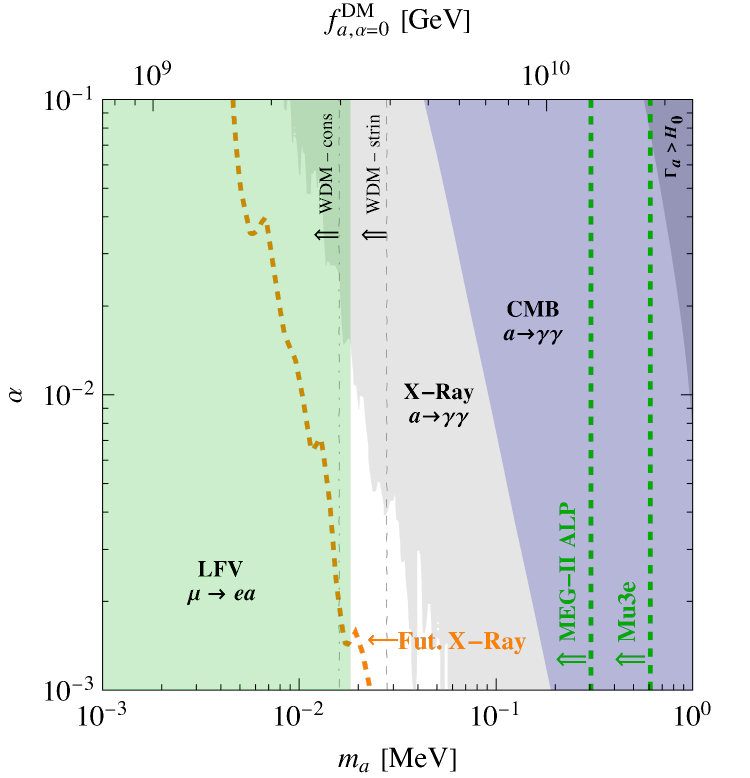}\hfill
		\includegraphics[width=0.47\textwidth]{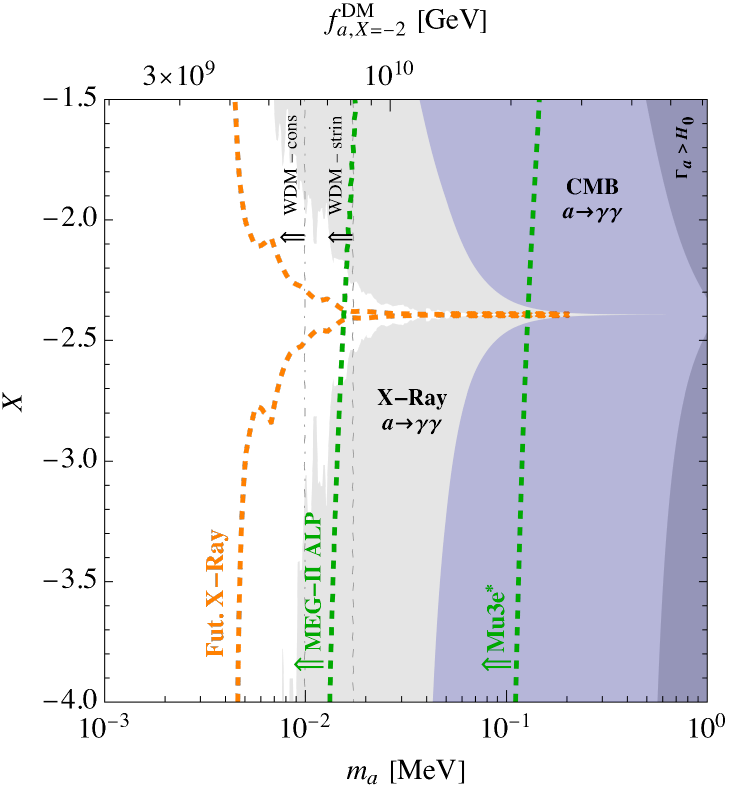}
	\caption{ Allowed parameter space for DM freeze-in through LFV decays. The decay constant $f_a$ (top x-axis) is determined by requiring that the DM abundance today is produced through freeze-in (see Sec.~\ref{sec:production} for details), once the ALP mass $m_a$ (bottom x-axis) and the remaining parameter $\alpha$ or $X$ (y-axis) is fixed (we choose the reference values $\alpha = 0$ and $X=-2$). The {\bf dark blue shaded}, {\bf blue shaded} and {\bf gray shaded} regions are excluded by the DM lifetime, CMB and X-ray constraints on decaying DM, respectively.  The reach of future X-rays searches is shown by {\bf dashed orange} lines  (see Appendix~\ref{sec:decayDM}).  Conservative (stringent) constraints on WDM requiring $m_{\rm WDM} \gtrsim 3.5 (5.3) \, \keV$ are recasted following Refs.~\cite{Ballesteros:2020adh,DEramo:2020gpr,Decant:2021mhj} and shown as {\bf dotted-dashed}  ({\bf dashed})  gray lines. The present bound from searches for $\mu \to e a$ (90\% CL) are shown as {\bf green shaded} regions, while the prospects for future proposed searches at MEG II~\cite{Jho:2022snj} and Mu3e~\cite{Calibbi:2020jvd,Perrevoort:2018ttp} are shown as  {\bf dashed green} lines. {\bf Left:} In the $\mu e$-scenario of Eq.~\eqref{eq:rh-scenario} flavor-diagonal couplings are suppressed by small $\alpha$, thus relaxing the X-rays constraints. Future X-ray searches will probe most of parameter space down to $\alpha\sim 10^{-3}$. The present bound on the LFV right-handed coupling of the ALP to muons~\cite{Jodidio:1986mz} sets already a strong lower bound on the ALP mass, and proposed searches at MEG II and Mu3e will probe almost the entire allowed parameter space independent of  $\alpha$.  {\bf Right:} In the PMNS scenario of Eq.~\eqref{eq:lh-scenario} the PQ charge $X$ is taken as a free parameter.  The WDM bounds assume that DM is produced solely from $\tau$-decays. The present bound on LFV left-handed couplings~\cite{TWIST:2014ymv} is too weak to appear in the plot, while future proposed searches at MEG II~\cite{Jho:2022snj} will be able to probe a large amount of parameter space, except a very fine-tuned region around $X\approx 2.4$. The reach of Mu3e for left-handed ALP couplings assumes that new calibration methods will be implemented in order to reduce systematic uncertainties (see Ref.~\cite{Calibbi:2020jvd} for a discussion). \label{fig:money}}
\end{center}
\end{figure*}

\section{Results}
We are now ready to summarize our results. Putting Eq.~\eqref{eq:lifetime} together with  Eq.~\eqref{eq:relic} and the lower bound on the DM mass in Eq.~\eqref{eq:WDM}, we rewrite the axion lifetime, assuming its abundance today comes dominantly from freeze-in of LFV decays, as 
\begin{widetext}
\begin{equation}
\tau_a^{\gamma\gamma}\approx 10^{26}\text{ sec}\left(\frac{C_{\ell_i \ell_j}}{C_{\ell_i \ell_i}}\right)^2\left(\frac{10\text{ keV}}{m_a}\right)^6\left( \frac{m_{\ell_i}}{m_e} \right)^{ 4}\left(\frac{75}{g_*(m_{\ell_j})}\right)^{3/2}\left(\frac{m_{\ell_j}}{ m_{\tau}}\right)\left(\frac{0.12}{\Omega_{\rm DM} h^2}\right)\, , 
\end{equation}
\end{widetext}
where $\ell_i$ ($\ell_j$) is the lighter (heavier) lepton.
This equation shows that generically a hierarchy between the LFV and flavor-diagonal couplings is \emph{necessary} to sufficiently suppress the axion diphoton width, in order to satisfy the stringent constraints from X-rays searches shown in Eq.~\eqref{eq:Xray_bound}. Since the parametric dependence of the freeze-in production through muon or tau LFV decays is very similar up to $\mathcal{O}(1)$ numerical factors, we can estimate the required hierarchy assuming that the axion freeze-in reproduces the total DM abundance today:
\begin{equation}
\frac{C_{\ell_i \ell_i}}{C_{\ell_i \ell_j}}\lesssim 0.1\left(\frac{10^{28}\text{ sec}}{\tau_a^{\gamma\gamma}}\right)^{1/2}\!\!\left(\frac{10\text{ keV}}{m_a}\right)^3\left( \frac{m_{\ell_i}}{m_e} \right)^{ 2}\, , \label{eq:hierarchy}
\end{equation}
where in the two-flavor scenario we have $C_{\ell_i \ell_i} / C_{\ell_i \ell_j} = \tan\alpha$.
This result makes it evident that a dominantly off-diagonal flavor texture of the axion couplings is required to open up the freeze-in DM parameter space for an axion coupled to electrons with $m_a\gtrsim10 \text{ keV}$. This feature would be strongly exacerbated by more stringent lower bounds on the DM mass from Ly$\alpha$ observations. For instance setting the most stringent present constraint $m_{\rm WDM}^{\rm min} \approx 5.3 \, \keV$ would already result in the upper bound  $m_a\gtrsim 17.4 \, \keV$ requiring a hierarchy $C_{\ell_i \ell_i}/C_{\ell_i \ell_j}\lesssim 10^{-2}$. Future X-rays searches are supposed to probe axion lifetimes around $10^{29}-10^{30}\text{ sec}$, requiring an even stronger suppression of the flavor-diagonal couplings. 

Interestingly, future LFV experiments can probe this axion production mechanism directly. In particular future proposals testing muon LFV decays will be able to probe $f_a/C_{\mu e}$ up to $10^{10}\text{ GeV}$, where the expected sensitivity and the best experimental strategy strongly  depends on the chiral structure of the axion couplings~\cite{Calibbi:2020jvd, Jho:2022snj,Perrevoort:2018ttp}. This sensitivity is essentially independent on the axion mass, which is well below the experimental resolution in the mass range of interest and can be taken to be zero for all the experimental purposes.  From Eq.~\eqref{eq:relic}, we conclude that the expected reach would be able to probe axion masses up to 100 keV. The remaining parameter space for axion mass between 100 keV and 1 MeV requires extreme hierarchies between the flavor-diagonal and LFV axion coupling as shown by Eq.~\eqref{eq:hierarchy}. 

We summarize our results in Fig.~\ref{fig:money}, which shows the allowed parameter space in the  $\mu e$- and PMNS scenarios. The axion decay constant $f_a$ has been fixed to reproduce the observed relic density, solving the Boltzmann equation numerically (the results are in very good agreement with the analytical approximations in Eq.~\eqref{eq:relic}). For small values of $\alpha$, the required value of $f_a$ is essentially independent of $\alpha$ and shown in the upper axis in the left panel. In the PMNS scenario the rotation angles are large, so that a sufficient suppression of the decay rate to photons can be obtained only in a tiny fine-tuned region of parameter space where $X \approx 2.6$, leading to an accidental cancellation in the axion-photon couplings.  

Finally we comment on other two-flavor cases where the LFV axion is coupled solely to taus and muons or taus and electrons. In the $\tau \mu$-scenario the suppression of the photon coupling by the muon mass is sufficient in order to allow for diagonal couplings of the same order of the LFV ones, giving a largely unconstrained parameter space. However, the expected sensitivity from searches for LFV tau  decays into axions at Belle-II only probes decay constants $f_a/C_{\tau \mu}$ up to $10^{8}\text{ GeV}$~\cite{Calibbi:2020jvd,Guadagnoli:2021fcj, Cheung:2021mol, Tenchini:2020njf},  which is unfortunately more than an order of magnitude too low to probe axion freeze-in (for prospects at a future muon collider see Ref.~\cite{Haghighat:2021djz}). This leaves only future X-ray telescopes to probe this scenario. The $\tau e$-scenario is very similar to the $\mu e$-scenario, except that the WDM bound is slightly relaxed and again the experimental prospects on tau LFV decays are too weak to probe the allowed parameter space. 
\section{Conclusions and Outlook}
We have proposed the possibility to produce axion DM from LFV decays of SM leptons. This gives rise to several simple scenarios, depending on the LFV transition and the relative size of diagonal and off-diagonal axion couplings to leptons, which essentially control axion decay and production, respectively. This ratio can be unity for tau decays to muons, while for tau or muon decays to electrons the stringent constraints from decaying DM together with the lower bound on the axion DM mass from WDM  constraints require some amount of suppression of diagonal couplings to electrons. Allowing for UV-sensitive scenarios where the DM abundance receives sizable contributions from both freeze-in and misalignment, the WDM constraint can be softened substantially opening up the allowed parameter space of these models. 

The allowed parameter space of all scenarios will be further tested by future X-ray telescopes, and in some cases almost entirely probed by precision measurements of rare muon decays. In particular,  the LFV muon decay width required for freeze-in sets a definitive target for proposed LFV experiments such as Mu3e~\cite{Calibbi:2020jvd,Perrevoort:2018ttp} and MEG II~\cite{Jho:2022snj}. These experiments have the unique opportunity of probing directly the very same decay that has produced axion DM in the early Universe. 

In the same spirit of this paper, it would be interesting to consider scenarios with flavor-violating axion couplings to quarks, where the $m_a^2/m_{\pi,K}^2$ suppression of the effective photon coupling would strongly  improve DM stability. 

\section*{Acknowledgments}
We thank Andrea Caputo for discussions on future X-rays searches, Joerg Jaeckel, Andrea Tesi, Michele Redi and Lorenzo Ubaldi for discussions on non-thermal production mechanisms and Ann-Kathrin Perrevoort for discussions on the expected reach of Mu3e and comments on the manuscript. 
We also thank Nick Rodd for feedback on the draft. This project has received support from the European Union's Horizon 2020 research and innovation programme under the Marie Sklodowska-Curie grant agreement No 860881-HIDDeN. This work is partially supported by project C3b of the DFG-funded Collaborative Research Center TRR257 ``Particle Physics Phenomenology after the Higgs Discovery". 

\appendix

\begin{figure*}[t]
\begin{center}
	\includegraphics[width=0.7\textwidth]{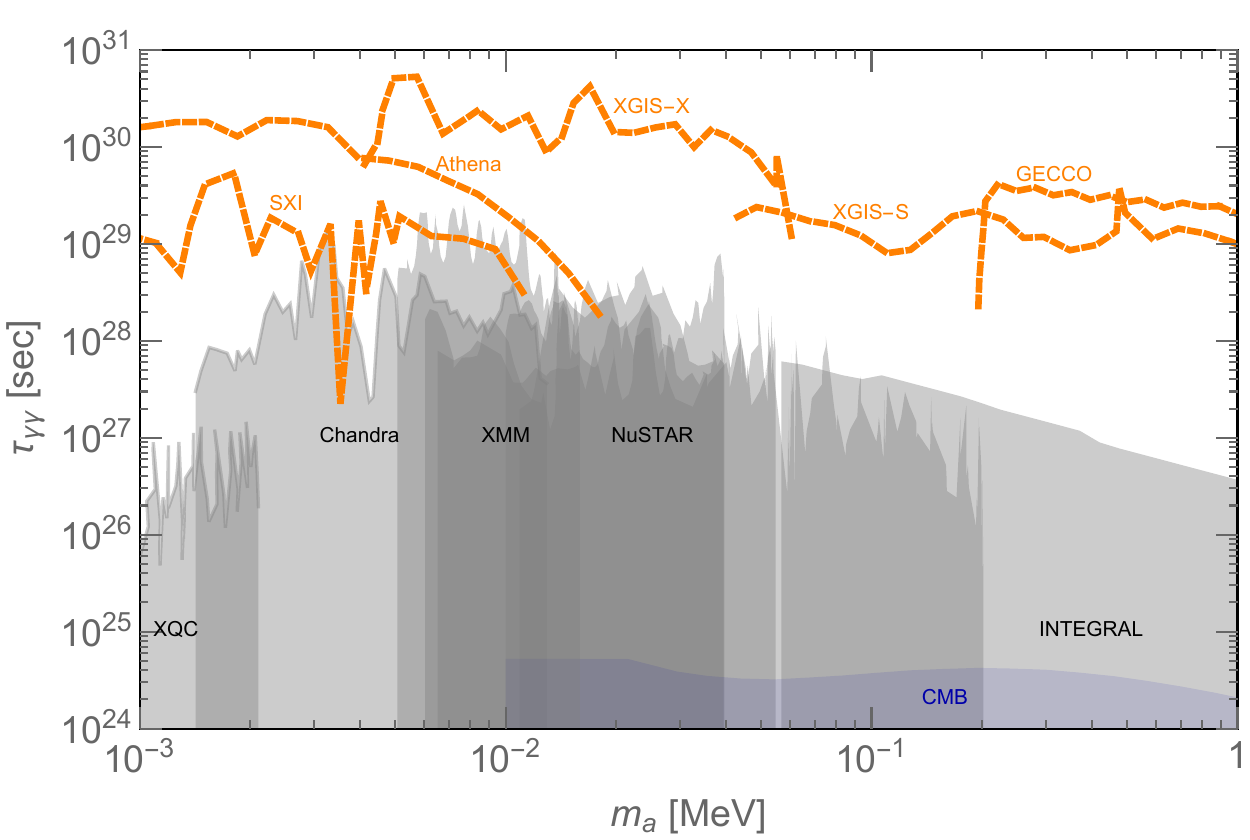}
	\caption{Summary of the present and future constraints on decaying dark matter in the keV-MeV mass range. The {\bf blue shaded} region is excluded by CMB constraints~\cite{Slatyer:2016qyl}. The {\bf gray shaded} regions are excluded by different X-rays and low energy gamma rays line searches: XQC~\cite{Boyarsky:2006hr,XQC:2015mwy}, Chandra~\cite{Horiuchi:2013noa,Watson:2011dw}, Newton-XMM~\cite{Foster:2022ajl}, NuStar~\cite{Perez:2016tcq,Roach:2019ctw,Ng:2019gch,Roach:2022lgo}, INTEGRAL~\cite{Laha:2020ivk}. We also show in {\bf dashed orange} the optimistic projections for future telescopes: GECCO~\cite{Coogan:2021rez}, the THESEUS mission composed by XGIS-S, XGIS-X and SXI~\cite{Thorpe-Morgan:2020rwc} and Athena~\cite{Neronov:2015kca,Dekker:2021bos,Ando:2021fhj}.}\label{fig:decayDM}
\end{center}
\end{figure*}

\section{Constraints on Decaying Dark Matter}\label{sec:decayDM}
We summarize here the different constraints on decaying dark matter into diphotons in the keV-MeV mass range, which are the ones of interest for our study. Our recollection is shown in Fig.~\ref{fig:decayDM}.

First we show in light blue the cosmological constraints on decaying DM which originate from the high energy photons injected in the photon-baryon fluid. These high energy particles 
induce an electromagnetic cascade, causing atomic ionizations, excitations and heating of the intergalactic medium~\cite{Chen:2003gz,Padmanabhan:2005es, Cirelli:2009bb, Galli:2009zc, Slatyer:2016qyl}, leaving imprints in the CMB anisotropies and leading to distortions of the CMB black body spectrum~\cite{Chluba:2011hw,Bolliet:2020ofj}. CMB anisotropies constrain axion diphoton lifetimes of order $10^{24}\text{ sec}$ for axion masses above 10 keV (see e.g. Ref.~\cite{Slatyer:2016qyl}). 

In the whole keV- MeV mass range, X-rays and low energy gamma-rays searches set stronger constraints than cosmology. In Fig.~\ref{fig:decayDM} we show bounds from the XQC sounding rocket~\cite{Boyarsky:2006hr,XQC:2015mwy} which sets the most stringent limit in a tiny mass region around $m_a\simeq 1\text{ keV}$. At higher masses, many searches have been performed since the 3.5 keV line excess was first found in Ref.~\cite{Bulbul:2014sua} analyzing the XMM-Newton observations in a stack of galaxy clusters.  We show the current best limits set by Chandra's observations of the M31 (Andromeda) galaxy~\cite{Watson:2011dw, Horiuchi:2013noa} and an updated analysis of the XMM-Newton dataset~\cite{Foster:2022ajl}.  At higher masses, the different observations of the \mbox{NuSTAR} telescope~\cite{Perez:2016tcq,Roach:2019ctw,Ng:2019gch,Roach:2022lgo} set the most stringent constraints on decaying DM. Between 100 keV and 1 MeV the most stringent constraint comes from  INTEGRAL full emission profile~\cite{Bouchet:2011fn} analyzed in Ref.~\cite{Laha:2020ivk}. It is worth noticing that this recent analysis got a less stringent constraint compared to Ref.~\cite{Essig:2013goa}, which was using  INTEGRAL data correlated with a specific emission template. 

We also collect in Fig.~\ref{fig:decayDM} the expected sensitivities of different future telescopes.  At high masses we show the projected sensivity of GECCO obtained in Ref.~\cite{Coogan:2021rez}. At lower masses an improved sensitivity is expected from the forthcoming THESEUS mission covering a wide energy range between 1 and 350 keV~\cite{Thorpe-Morgan:2020rwc}. At lower energies future X-ray missions like e-ROSITA and especially Athena are expected to probe dark matter lifetimes exceeding $10^{30}\text{sec}$~\cite{Neronov:2015kca,Dekker:2021bos,Ando:2021fhj} down to a few keV. The reach around 1 keV mass region will also be extended by Micro-X observations~\cite{Adams:2019nbz}.

\bibliographystyle{JHEP}
\bibliography{PRSZ}

\end{document}